\documentclass[prd,showpacs,amsmath,amssymb,twocolumn]{revtex4}
\usepackage{graphicx,color,slashbox}
\begin{document}


\title{
The Hidden Charm Decay of $X(3872), Y(3940)$ and Final State
Interaction Effects}

\author{Xiang Liu}\email{xiangliu@pku.edu.cn}\author{Bo
Zhang}\author{Shi-Lin Zhu}

\vspace*{1.0cm}

\affiliation{Department of physics, Peking University, Beijing,
100871, China}

\vspace*{1.0cm}

\date{\today}
\begin{abstract}

We investigate whether the final state interaction (FSI) effect
plays a significant role in the large hidden charm decay width of
X(3872) and Y(3940) using a model. Our numerical result suggests
(1) the FSI contribution to $X(3872)\to J/\psi\rho $ is tiny; (2)
$\Gamma[ Y(3940)\to D\bar{D}^{*}+\text{h.c.}\to J/\psi\omega ]$
from FSI is around several keV, far less than Belle's experimental
value 7 MeV.

\end{abstract}

 \maketitle

\newpage

\section{Introduction}\label{sec1}

The underlying structure of $X(3872)$ is still very controversial.
It was discovered by Belle collaboration \cite{belle-3872} and
confirmed by Babar \cite{babar-3872}, CDF \cite{CDF-3872} and D0
\cite{D0-3872} collaborations. Recently, Belle collaboration
reported a new decay mode $X(3872)\to D^{0}\bar{D}^{0}\pi^{0}$
\cite{DDpi-3872}. The mass of $X(3872)$ from various experiments
reads \cite{belle-3872,babar-3872,CDF-3872,D0-3872,DDpi-3872}
\begin{eqnarray}
 M_{_{X(3872)}}= \left \{ \begin{array}{lc}
3875\pm0.7^{+1.2}_{-2.0}\; {\text MeV}/{\text c}^2&{\text Belle}
 \\
 3871.8 \pm 3.1 \pm 3.0\; {\text MeV}/{\text c}^2 &
 {\text D0}
 \\
 3871.3 \pm 0.7\pm 0.4 \; {\text MeV}/{\text c}^2&
  {\text CDF}
  \\
  3873.4 \pm 1.4 \; {\text MeV}/{\text c}^2&
  {\text BaBar}
  \\3872.0 \pm 0.6 \pm 0.5 \; {\text MeV}/{\text c}^2&
  {\text Belle}.
  \end{array}\right.
\end{eqnarray}
The available experimental information indicates $J^{PC}=1^{++}$ for
$X(3872)$ \cite{angular-3872}. Theoretical interpretations of
$X(3872)$ include a charmonium state \cite{charmonium}, a molecular
state \cite{mole}, or the mixture of charmonium with molecular state
\cite{mix}.

At present the observed decay modes of $X(3872)$ include
$J/\psi\pi^{+}\pi^{-}$ \cite{belle-3872}, $\gamma J/\psi$
\cite{jpsi-gamma}, $J/\psi\pi^{+}\pi^{-}\pi^{0}$ \cite{jpsi-gamma}
and $D^{0}\bar{D}^{0}\pi^{0}$ \cite{DDpi-3872}. The dipion in
$J/\psi\pi^{+}\pi^{-}$ seems to originate from $\rho\to
\pi^{+}\pi^{-}$ because the peak of the dipion invariant mass
spectrum locates around 775 MeV. $J/\psi\pi^{+}\pi^{-}\pi^{0}$
comes from the sub-threshold decay $X(3872)\to J/\psi\omega$
\cite{jpsi-gamma}. Meanwhile the ratio of $B(X(3872)\to J/\psi
\pi^{+}\pi^{-}\pi^{0})$ to $B(X(3872)\to J/\psi\pi^{+}\pi^{-})$
given by experiment is $1.0\pm 0.4({\rm stat})\pm 0.3({\rm syst})$
\cite{jpsi-gamma}. Recently Belle experiment indicated
$B(X(3872)\to D^{0}\bar{D}^{0}\pi^{0}K^{+})=9.4^{+3.6}_{-4.3}
B(X(3872)\to J/\psi\pi^{+}\pi^{-}K^{+})$ \cite{DDpi-3872}. Based
on the above experimental data, one concludes (1) the
$D^{0}\bar{D}^{*0}$ is the dominant decay of $X(3872)$; (2) the
isospin violating mode $X(3872)\to J/\psi\rho\to
J/\psi\pi^{+}\pi^{-}$ is not suppressed, compared with the isospin
conserving mode $X(3872)\to J/\psi\omega\to J/\psi
\pi^{+}\pi^{-}\pi^{0}$.

The Final State Interaction (FSI) effect sometimes plays a crucial
role in many processes \cite{FSI}. In this work, we study if the
hidden charm decay $J/\psi\rho(\pi^{+}\pi^{-})$ mainly arises from
the FSI effect of $X(3872)\to \bar{D}^{*0}D^{0}+{\rm{h.c.}}$.
$X(3872)$ decays to $D^{0}\bar{D}^{*0}+\text{h.c.}$ but not
$D^{+}D^{*-} +\text{h.c.}$ because $D^{0}+\bar{D}^{*0}= 3871.3
\;\text{MeV}<M_{_{X(3872)}}$ and $D^{+}+D^{*-}=3879.3\;\text{MeV}
>M_{_{X(3872)}}$. Thus the isospin violating process $X(3872)\to
J/\psi\rho$ can occur via the $\bar{D}^{*0}D^{0}+{\rm{h.c.}}$
re-scattering effect.

This paper is organized as follows. We present the formulation about
$X(3872)\to D^{0}\bar{D}^{*0}+{\rm{h.c.}}\to J/\psi\rho$ in Section
\ref{sec2}. Then we present our numerical results. The last section
is a short discussion.

\section{Formalism}\label{sec2}

\begin{figure}[htb]
\begin{center}
\scalebox{0.5}{\includegraphics{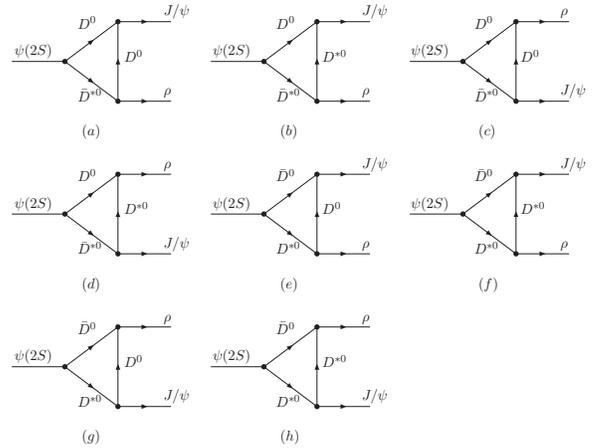}}
\end{center}
\caption{The diagrams for $X(3872)\to
\bar{D}^{*0}D^{0}+{\rm{h.c.}}\to J/\psi\rho$.}\label{FSI}
\end{figure}

The Feynman diagrams for the $X(3872)\to J/\psi\rho$ through
$\bar{D}^{*0}D^{0}+{\rm{h.c.}}$ re-scattering are depicted in Fig.
\ref{FSI}.

In Refs. \cite{lagrangian-jpsi,lagrangian-hl,Casalbuoni}, the
effective Lagrangians, which are relevant to the present
calculation, are constructed based on the chiral symmetry and
heavy quark symmetry:
\begin{eqnarray}
\mathcal{L}&=&g_{X}X^{\mu}[D^{0}\bar{D}^{*0}_{\mu}-\bar{D}^{0}{D}^{*0}_{\mu}
]\nonumber\\&&+i g_{_{J/\psi \mathcal{D}\mathcal{D}}}^{} \psi_\mu
\left(
\partial^\mu \mathcal{D} {\mathcal{D}}^{\dagger} - \mathcal{D}
\partial^\mu {\mathcal{D}}^{\dagger}
\right)\nonumber\\&& -g_{_{J/\psi \mathcal{D}^* \mathcal{D}}}^{}
\varepsilon^{\mu\nu\alpha\beta}
\partial_\mu \psi_\nu \left(
\partial_\alpha \mathcal{D}^*_\beta {\mathcal{D}}^{\dagger} + \mathcal{D} \partial_\alpha {\mathcal{D}}^{*\dagger}_\beta
\right)\nonumber\\
&&-i g_{_{J/\psi \mathcal{D}^* \mathcal{D}^*}}^{} \Bigl\{ \psi^\mu
\left(
\partial_\mu \mathcal{D}^{*\nu} {\mathcal{D}}_\nu^{*\dagger} -
\mathcal{D}^{*\nu}
\partial_\mu {\mathcal{D}}_\nu^{*\dagger} \right)
\nonumber\\&&+ \left( \partial_\mu \psi_\nu \mathcal{D}^{*\nu} -
\psi_\nu
\partial_\mu \mathcal{D}^{*\nu} \right) {\mathcal{D}}^{*\mu\dagger}  \mbox{}  + \mathcal{D}^{*\mu}
\big( \psi^\nu
\partial_\mu {\mathcal{D}}^{*\dagger}_{\nu} \nonumber\\&&- \partial_\mu \psi_\nu {\mathcal{D}}^{*\nu\dagger}
\big) \Bigr\}
+\Big\{-ig_{_{\mathcal{D}\mathcal{D}\mathbb{V}}}\mathcal{D}_{i}^{\dagger}{\stackrel{\leftrightarrow}{\partial}}
_{\mu}\mathcal{D}^{j}(\mathbb{V}^{\mu})^{i}_{j}\nonumber\\&&
-2f_{_{\mathcal{D^{*}}\mathcal{D}\mathbb{V}}}\varepsilon_{\mu\nu\alpha\beta}(\partial^{\mu}\mathbb{V}^{\nu})
^{i}_{j}(\mathcal{D}_{i}^{\dagger}
{\stackrel{\leftrightarrow}{\partial}}^{\alpha}\mathcal{D^{*}}^{\beta
j}-\mathcal{D^{*}}_{i}^{\beta
\dagger}{\stackrel{\leftrightarrow}{\partial}}^{\alpha}\mathcal{D}^{j})\nonumber\\
&&+ig_{_{\mathcal{D^{*}}\mathcal{D^{*}}\mathbb{V}}}\mathcal{D^{*}}_{i}^{\nu
\dagger}{\stackrel{\leftrightarrow}{\partial}}_{\mu}\mathcal{D^{*}}_{\nu}^{j}(\mathbb{V}^{\mu})^{i}_{j}\nonumber\\&&+
4if_{_{\mathcal{D^{*}}\mathcal{D^{*}}\mathbb{V}}}\mathcal{D^{*}}_{i\mu}^{\dagger}(\partial^{\mu}\mathbb{V}^{\nu}-\partial^{\nu}
\mathbb{V}^{\mu})^{i}_{j}
\mathcal{D^{*}}_{\nu}^{j}\Big\},\label{lagrangian}
\end{eqnarray}
where $\mathcal{D}$ and $\mathcal{D^*}$ are the pseudoscalar and
vector heavy mesons respectively, i.e.
$\mathcal{D^{(*)}}$=(($\bar{D}^{0})^{(*)}$, $(D^{-})^{(*)}$,
$(D_{s}^{-})^{(*)}$). The values of the coupling constants will be
given in the following section. $\mathbb{V}$ denotes the nonet
vector meson matrices
\begin{eqnarray}
\mathbb{V}&=&\left(\begin{array}{ccc}
\frac{\rho^{0}}{\sqrt{2}}+\frac{\omega}{\sqrt{2}}&\rho^{+}&K^{*+}\\
\rho^{-}&-\frac{\rho^{0}}{\sqrt{2}}+\frac{\omega}{\sqrt{2}}&
K^{*0}\\
K^{*-} &\bar{K}^{*0}&\phi
\end{array}\right).
\end{eqnarray}

By the Cutkosky rule, the absorptive part of Fig. 1 (a) which comes
from the re-scattering process of $X(3872)\to
D^{0}(p_{1})+\bar{D}^{*0}(p_{2})\to J/\psi(p_{3})+\rho(p_{4})$ is
written as
\begin{eqnarray}
\textbf{Abs}(a)&=&\frac{1}{2}\int\frac{d^{3}p_{1}}{(2\pi)^{3}2E_{1}}\frac{d^{3}p_{2}}{(2\pi)^{3}2E_{2}}
\nonumber\\&&\times(2\pi)^{4}\delta^{4}(m_{_X}-p_{1}-p_{2})
[ig_{X}\varepsilon_{\xi}]\nonumber\\&&\times\Big[-g_{_{J/\psi
DD}}i(p_{1}-q)\cdot
\varepsilon_{J/\psi}\Big]\nonumber\\
&&\times\Big[-\frac{2}{\sqrt{2}}i\;f_{_{D^{*}DV}}\epsilon_{\mu\nu\alpha\beta}ip_{4}^{\mu}\varepsilon_{\rho}^{\nu}(iq^{\alpha}
+ip_{2}^{\alpha})\Big]\nonumber\\&&\times\bigg[-g^{\xi\beta}+\frac{p_{2}^{\xi}p_{2}^{\beta}}{m_{2}^{2}}\bigg]\bigg[\frac{i}{q^2
-m_{D}^{2}}\bigg]\mathcal{F}^{2}(m_{D},q^2)\nonumber\\
&=&\int
d\Omega\frac{|\mathbf{p}_{1}|}{32\pi^{2}m_{_X}}[\sqrt{2}g_{X}g_{_{J/\psi
DD}}f_{_{D^{*}DV}}]\nonumber\\&&\times[(2p_{1}-p_{3})\cdot
\varepsilon_{J/\psi}]
\epsilon_{\mu\nu\alpha\beta}p_{4}^{\mu}\varepsilon_{\rho}^{\nu}(p_{3}^{\alpha}+p_{2}^{\alpha}-p_{1}^{\alpha})
\nonumber\\&&\times\bigg[-\varepsilon^{\beta}+p_{2}^{\beta}\frac{p_{2}\cdot
\varepsilon}{m_{2}^{2}}\bigg]
\frac{\mathcal{F}^{2}(m_{D},q^2)}{q^{2}-m_{D}^{2}}
\end{eqnarray}
with
$q^{2}=m_{1}^{2}+m_{3}^{2}-2E_{1}E_{3}+2|\mathbf{p}_{1}||\mathbf{p}_{3}|\cos\theta$,
where $\mathcal{F}^{2}(m_{i},q^2)$ etc denotes the form factors
which compensate the off-shell effects of mesons at the vertices
and are written as
\begin{eqnarray}
\mathcal{F}^{2}(m_{i},q^2)=\bigg(\frac{\Lambda^{2}-m_{i}^2
}{\Lambda^{2}-q^{2}}\bigg)^2,
\end{eqnarray}
where $\Lambda$ is a phenomenological parameter. As $q^2\to 0$ the
form factor becomes a number. If $\Lambda\gg m_{i}$, it becomes
unity. As $q^2\rightarrow\infty$, the form factor approaches to
zero. As the distance becomes very small, the inner structure
would manifest itself and the whole picture of hadron interaction
is no longer valid. Hence the form factor vanishes and plays a
role to cut off the end effect. The expression of $\Lambda$ is
\cite{HY-Chen}
\begin{eqnarray}
\Lambda(m_{i})=m_{i}+\alpha \Lambda_{QCD},
\end{eqnarray}
where $m_{i}$ denotes the mass of exchanged meson.
$\Lambda_{QCD}=220$ MeV. $\alpha$ is a phenomenological parameter.

Similarly we obtain the absorptive contributions from Fig.
\ref{FSI} (b)-(d) respectively.
\begin{eqnarray}
\textbf{Abs}(b)&=&\frac{1}{2}\int\frac{d^{3}p_{1}}{(2\pi)^{3}2E_{1}}\frac{d^{3}p_{2}}{(2\pi)^{3}2E_{2}}\nonumber\\&&
\times(2\pi)^{4}\delta^{4}(m_{_X}-p_{1}-p_{2})
[ig_{X}\varepsilon_{\xi}]\nonumber\\&& \times\Big[i\;g_{_{J/\psi
DD^{*}}}\epsilon_{\mu\nu\kappa\sigma}\varepsilon_{J/\psi}^{\mu}(-i)p_{1}^{\nu}(-i)q^{\sigma}\Big]\nonumber\\&&\times
\bigg\{-\frac{g_{_{D^{*}D^{*}V}}}{\sqrt{2}}
i(q+p_{2})\cdot\epsilon_{\rho}g_{\alpha\beta}\nonumber\\&&
-\frac{4f_{_{D^{*}D^{*}V}}}{\sqrt{2}}\Big[ip_{4\beta}{\epsilon_{\rho}}_{\alpha}-i
{\epsilon_{\rho}}_{\beta}p_{4\alpha}\bigg]\bigg\}\nonumber\\&&
\times\bigg[-g^{\kappa\beta}+\frac{p_{2}^{\kappa}p_{2}^{\beta}}{m_{2}^{2}}\bigg]
\bigg[-g^{\xi\alpha}+\frac{q^{\xi}q^{\alpha}}{m_{D^{*}}^{2}}\bigg]\nonumber\\&&\times\bigg[\frac{i}{q^2
-m_{D^{*}}^{2}}\bigg]\bigg(\frac{\Lambda^{2}-m_{D^{*}}^2
}{\Lambda^{2}-q^{2}}\bigg)^2,
\end{eqnarray}
\begin{eqnarray}
\textbf{Abs}(c)&=&\frac{1}{2}\int\frac{d^{3}p_{1}}{(2\pi)^{3}2E_{1}}\frac{d^{3}p_{2}}{(2\pi)^{3}2E_{2}}
\nonumber\\&& \times(2\pi)^{4}\delta^{4}(m_{_X}-p_{1}-p_{2})
[ig_{X}\varepsilon_{\xi}]\nonumber\\&&
\times\bigg[\frac{g_{_{DDV}}}{\sqrt{2}}i(q-p_{1})\cdot\varepsilon_{\rho}\bigg]\nonumber\\&&
\times \Big[ig_{_{J/\psi DD^{*}}}
\epsilon_{\mu\nu\alpha\beta}\varepsilon_{J/\psi}^{\mu}iq^{\nu}(-i)p_{2}^{\beta}\Big]\nonumber\\&&
\times\bigg[-g^{\xi\alpha}+\frac{p_{2}^{\xi}p_{2}^{\alpha}}
{m_{2}^{2}}\bigg]\bigg[\frac{i}{q^2
-m_{D}^{2}}\bigg]\bigg(\frac{\Lambda^{2}-m_{D}^2
}{\Lambda^{2}-q^{2}}\bigg)^2,\nonumber\\
\end{eqnarray}
\begin{eqnarray}
\textbf{Abs}(d)&=&\frac{1}{2}\int\frac{d^{3}p_{1}}{(2\pi)^{3}2E_{1}}\frac{d^{3}p_{2}}{(2\pi)^{3}2E_{2}}\nonumber\\&&
\times(2\pi)^{4}\delta^{4}(m_{_X}-p_{1}-p_{2})
[ig_{X}\varepsilon_{\xi}]\nonumber\\&&
\times\bigg[-\frac{2}{\sqrt{2}}if_{_{D^{*}DV}}\epsilon_{\mu\nu\alpha\beta}ip_{3}^{\mu}\varepsilon_{\rho}^{\nu}
i(q^{\alpha}-p_{1}^{\alpha})\bigg]\nonumber\\&&\times\Big\{-g_{_{J/\psi
D^{*}D^{*}}}\Big[iq^{\kappa}\varepsilon_{J/\psi}^{\sigma}+ip_{2}^{\sigma}\varepsilon_{J/\psi}^{\kappa}\nonumber\\&&
\times+i(p_{2}+q)\cdot\varepsilon_{J/\psi}g^{\kappa\sigma} \Big]
\Big\}\bigg[-g^{\xi}_{\kappa}+\frac{{p_{2}}_{\kappa}p_{2}^{\xi}}{m_{2}^{2}}\bigg]
\nonumber\\&&\times\bigg[-g^{\beta}_{\sigma}+\frac{q_{\sigma}q^{\beta}}{m_{D^{*}}^{2}}\bigg]\bigg[\frac{i}{q^2
-m_{D^{*}}^{2}}\bigg]\bigg(\frac{\Lambda^{2}-m_{D^{*}}^2
}{\Lambda^{2}-q^{2}}\bigg)^2.\nonumber\\
\end{eqnarray}
The contributions from Fig. \ref{FSI} (e), (f), (g), (h) is the
same as that corresponding to Fig. \ref{FSI} (a), (b), (c), (d)
respectively.

The total amplitude of $X(3872)\to
D^{0}\bar{D}^{*0}+\bar{D}^{0}D^{*0}\to J/\psi\rho$ can be written
as
\begin{eqnarray}
\mathcal{M}&=&2[\textbf{Abs}(a)+\textbf{Abs}(b)+\textbf{Abs}(c)+\textbf{Abs}(d)],
\end{eqnarray}
where the pre-factor "2" comes from the consideration that the
contribution from $D^{0}\bar{D}^{*0}$ re-scattering is the same as
that from $\bar{D}^{0}D^{*0}$ re-scattering.

Because the $\rho$ meson is a broad resonance with
$\Gamma_{\rho}\sim 150$ MeV, the decay width of $X(3872)\to
D^{0}\bar{D}^{*0}+\bar{D}^{0}D^{*0}\to J/\psi\rho$ is written as
\begin{eqnarray}
\Gamma=\int^{(M_{_{X(3872)}}-m_{J/\psi})^{2}}_{0}{\rm{d}}s
f(s,m_{\rho},\Gamma_{\rho})\frac{|\mathbf{k}||\mathcal{M}(m_{\rho}\to
\sqrt{s})|^{2}}{24\pi M^{2}_{_{X(3872)}}},\nonumber
\end{eqnarray}
where the Breit-Wigner distribution function
$f(s,m_{\rho},\Gamma_{\rho})$ and the decay momentum
$|\mathbf{k}|$ are
\begin{eqnarray*}
&f(s,m_{\rho},\Gamma_{\rho})=\frac{1}{\pi}
\frac{m_{\rho}\Gamma_{\rho}}{(s-m_{\rho}^{2})^{2}+m_{\rho}^{2}\Gamma_{\rho}^{2}},\\
&|\mathbf{k}|=\frac{\sqrt{[M_{_{X(3872)}}^{2}-(\sqrt{s}+m_{J/\psi})^{2}][M_{_{X(3872)}}^{2}-(\sqrt{s}-m_{J/\psi})^{2}]}}{2M_{_{X(3872)}}}\;.
\end{eqnarray*}

\section{Numerical Results}\label{sec3}

The coupling constants related to our calculation include
\cite{HY-Chen}:
\begin{eqnarray*}
g_{_{DDV}}&=&g_{_{D^{*}D^{*}V}}=\frac{\beta
g_{_{V}}}{\sqrt{2}},\;\;\;f_{_{D^{*}DV}}=\frac{f_{_{D^{*}D^{*}V}}}{m_{_{D^*}}}=\frac{\lambda
g_{_{V}}}{\sqrt{2}},\nonumber\\
g_{_{V}}&=&\frac{m_{_{\rho}}}{f_{\pi}},
\end{eqnarray*}
where $f_{\pi}=132$ MeV, $g_{_{V}},\;\beta$ and $\lambda$ are
parameters in the effective chiral Lagrangian that describes the
interaction of heavy mesons with the low-momentum light vector
mesons \cite{Casalbuoni}. Following Ref. \cite{Isola}, we take
$g=0.59$, $\beta=0.9$ and $\lambda=0.56$. Based on the vector
meson dominance model and using $J/\psi$'s leptonic width, the
authors of Ref. \cite{Achasov} determined ${g_{_{J/\psi
\mathcal{D} \mathcal{D}}}^2}/{(4\pi)}=5$. As a consequence of the
spin symmetry in the heavy quark effective field theory,
$g_{_{J/\psi \mathcal{D}{\mathcal{D}}^{*}}}$ and $g_{_{J/\psi
\mathcal{D}^{*}{\mathcal{D}}^{*}}}$ satisfy the relations:
$g_{_{J/\psi \mathcal{D}{\mathcal{D}}^{*}}}={g_{_{J/\psi
\mathcal{D}{\mathcal{D}}}}}/{m_{_{D}}}$ and $g_{_{J/\psi
\mathcal{D}^{*}{\mathcal{D}}^{*}}}=g_{_{J/\psi
\mathcal{D}{\mathcal{D}}}}$ \cite{JPsi-relation}.

By fitting the upper limit of the total width of $X(3872)$ (2.3
MeV), one obtains the coupling constant $g_{X}$ in Eq.
(\ref{lagrangian})
\begin{eqnarray*}
g_{X}=\left\{\begin{array}{cc}
 2.2 \;\mathrm{GeV}, &\mathrm{for}\quad M_{_{X(3872)}}=3875.0\;\mathrm{MeV} ,\\
 2.5\;\mathrm{GeV},&\mathrm{for}\quad M_{_{X(3872)}}=3873.4\;\mathrm{MeV},\\
 3.1\;\mathrm{GeV},&\mathrm{for}\quad M_{_{X(3872)}}=3872.0\;\mathrm{MeV},
\end{array}\right.
\end{eqnarray*}
where we approximately take $\bar{D}^{*0}D^{0}$ as the dominant
decay mode of $X(3872)$ considering the experimental result
\cite{DDpi-3872}: $B(X(3872)\to
D^{0}\bar{D}^{0}\pi^{0}K^{+})=9.4^{+3.6}_{-4.3} B(X(3872)\to
J/\psi\pi^{+}\pi^{-}K^{+})$.

The value of $\alpha$ in the form factor usually is of order unity
\cite{HY-Chen}. In this work we take the range of $\alpha=0.5\sim
3$. The dependence of the branching ratio of $X(3872)\to
D^{0}\bar{D}^{*0}\to J/\psi\rho $ on $\alpha$ is presented in Fig.
2.

In Table \ref{numerical}, we list the typical values of the
branching ratio of $X(3872)\to D^{0}\bar{D}^{*0}+{\rm{h.c.}} \to
J/\psi\rho$ when we take several masses of $X(3872)$ from various
experiments and different $\alpha$.

\begin{figure}
\begin{center}
\scalebox{0.7}{\includegraphics{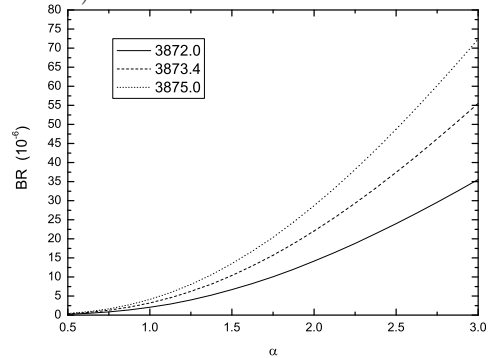}}
\end{center}
\caption{The dependence of decay width of $X(3872)\to
D^{0}\bar{D}^{*0}+{\rm{h.c.}}\to J/\psi\rho $ on $\alpha$.
}\label{fig1}
\end{figure}

\begin{widetext}

\begin{table}[h]
\begin{center}
\begin{tabular}{c|cccccccccc}
\hline \backslashbox{Mass (MeV)}{$\alpha$}& 0.5 & 1.0 & 1.5 & 2.0 &
2.5 & 3.0 \\ \hline\hline 3872.0 \cite{belle-3872}&$2.1\times
10^{-7}$&$2.0\times10^{-6}$ &
$6.6\times10^{-6}$&$1.4\times10^{-5}$&$2.4\times10^{-5}$&$3.6\times10^{-5}$
\\ \hline 3873.4
\cite{babar-3872}&$3.2\times10^{-7}$&$3.2\times10^{-6}$&$1.0\times10^{-5}$&$2.2\times10^{-5}$&$3.7\times10^{-5}$&$5.6\times10^{-5}$
&\\\hline
3875.0 \cite{DDpi-3872}&$4.2\times10^{-7}$&$4.1\times10^{-6}$&$1.3\times10^{-5}$&$2.9\times10^{-5}$&$4.9\times10^{-5}$&$7.2\times10^{-5}$&&&\\
 \hline\hline
\end{tabular}
\end{center}
\caption{The typical values of branching ratio of $X(3872)\to
D^{0}\bar{D}^{*0}+{\rm{h.c.}} \to J/\psi\rho$ for different values
of $M_{X(3872)}$ and $\alpha$.} \label{numerical}
\end{table}
\end{widetext}
\section{Discussion}\label{sec4}

Understanding the large $J/\psi\rho$ decay width of $X(3872)$ may
help reveal the nature of $X(3872)$. In this work, we study if the
large branching ratio of $X(3872)\to J/\psi\rho$ can be explained
by the $X(3872)\to D^{0}\bar{D}^{0*}+{\rm{h.c.}}$ re-scattering
effect. The numerical results indicate that $B(X(3872)\to
D^{0}\bar{D}^{0*}+{\rm{h.c.}}\to J/\psi \rho)$ is about
$10^{-5}\sim 10^{-7}$. Thus the large isospin violating
$X(3872)\to J/\psi\rho$ decay width can hardly be attributed to
the FSI effect of $X(3872)\to D^{0}\bar{D}^{0*}+{\rm{h.c.}}$. The
suppression from the phase space of $X(3872)\to
D^{0}\bar{D}^{0*}+{\rm{h.c.}}$ is huge because the experimental
values of $X(3872)$ mass is only barely above the $J/\psi+\rho$ or
$D^{*0}+D^{0}$ threshold, although the re-scattering effect is
obvious.

The reliable dynamical calculation of the hidden charm decay width
has been a challenging theoretical issue for decades. In Ref.
\cite{swanson}, the explicit $J/\psi\rho$ component is introduced
into the $X(3872)$ wave function in order to explain the large
$J/\psi\pi\pi$ decay width of $X(3872)$. In Ref. \cite{Yan-Kuang},
QCD multipole expansion technique was used to calculate hadronic
transitions such as $\psi(2S)\to J/\psi\pi\pi$ and
$\Upsilon(ns)\to \Upsilon(1s)\pi\pi$. If the main component of
$X(3872)$ is $c\bar{c}$ \cite{mix,suzuki}, the large $X(3872)\to
J/\psi\pi\pi$ decay might also be understood with this approach.

\begin{figure}
\begin{center}
\scalebox{0.7}{\includegraphics{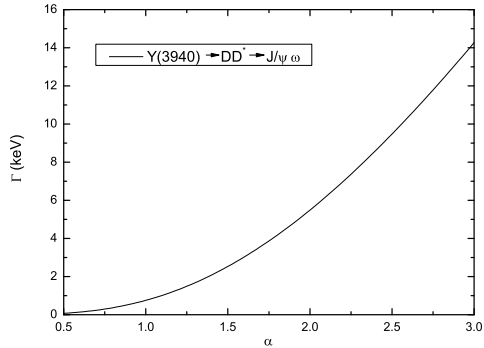}}
\end{center}
\caption{The dependence of the decay width of $Y(3940)\to
D\bar{D}^{*}\to J/\psi\omega$ on $\alpha$.}\label{result-3940}
\end{figure}

Last year, Belle reported a new charmonium state $Y(3940)$ in the
channel $B\to J/\psi\omega K$ and obtained $B(B\to Y(3940)+K)\cdot
B(Y(3940)\to \omega J/\psi)=(7.1\pm1.3\pm 3.1)\times 10^{-5}$. Its
mass and width are $3946\pm11(stat)\pm13(syst)$ MeV and
$87\pm22(stat)\pm 26(syst)$ MeV \cite{3940} respectively. In
particular, $\Gamma[Y(3940)\to J/\psi \omega] > 7$ MeV. Godfrey
suggested $Y(3940)$ as the $\chi_{c1}'$ state with quantum number
$2^{3}P_{1}$ \cite{Godfrey} and indicated that $Y(3940)\to
J/\psi\omega$ might come from the FSI effect of $Y(3940)\to
D\bar{D}^{*}+{\rm{h.c.}}$. In Fig. \ref{result-3940}, we present the
dependence of the width of $Y(3940)\to D\bar{D}^{*}+{\rm{h.c.}}\to
J/\psi\omega$ on $\alpha$\footnote{For the calculation of
$Y(3940)\to D\bar{D}^{*}\to J/\psi\omega$, we only replace relevant
masses and coupling constant in the formulas of $X(3872)\to
D^{0}\bar{D}^{*0}+{\rm{h.c.}} \to J/\psi\rho$ to make a rough
estimate. Other intermediate mesons for $Y(3940)\to J/\psi\omega$
are also allowed.}, where one takes $D\bar{D}^{*}$ as the dominant
decay mode of $Y(3940)$ as suggested in Ref. \cite{Godfrey}. The
order of magnitude of $\Gamma[Y(3940)\to D\bar{D}^{*}+{\rm{h.c.}}\to
J/\psi\omega]$ is keV, which is far less than Belle's data. Clearly
more experimental information on $Y(3940)$ will be very helpful.

\section*{Acknowledgments}
S.L.Z thanks Prof. K.T. Chao for helpful discussions. This project
was supported by the National Natural Science Foundation of China
under Grants 10375003, 10421503 and 10625521, Key Grant Project of
Chinese Ministry of Education (No. 305001).

\end{document}